# WHY DOES THE ENGINEERING MANAGER STILL EXIST IN AGILE SOFTWARE DEVELOPMENT?

## Ravi Kalluri

Assistant Professor of Project Management, College of Professional Studies, Northeastern University, San Jose, California, USA

### ABSTRACT

*Although Agile methodologies emphasise decentralised decision-making and team autonomy, engineering managers continue to be employed in Agile software organisations. This apparent paradox suggests that traditional managerial functions persist despite Agile's theoretical displacement of managerial hierarchy. This paper explores the persistence of engineering managers through a multidimensional framework encompassing historical context, theoretical tensions, organisational realities, empirical evidence, evolving managerial roles, and practical implications. A systematic literature review underpins our multifaceted analysis, supplemented by illustrative case studies. We conclude by proposing a conceptual model that reconciles Agile principles with managerial necessity, offering guidance for practitioners, researchers, and tool designers. Implications for leadership development, tool integration, and future research are discussed.*

## 1. INTRODUCTION

Agile software development frameworks—most notably Scrum, Extreme Programming (XP), and Kanban—were introduced in response to the limitations of traditional, plan-driven methodologies such as Waterfall (Beck et al., 2001; Dybå & Dingsøyr, 2008). By championing iterative delivery, close customer collaboration, and adaptive planning, the Agile Manifesto explicitly prioritised "individuals and interactions over processes and tools" (Beck et al., 2001, para. 1). Self-organizing teams lie at the heart of Agile ideology, empowered to make technical and process decisions without hierarchical oversight (Hoda et al., 2013). Early Agile literature and practitioner guides even suggested that traditional engineering managers were obsolete in Agile environments, as Scrum Masters and Product Owners could absorb managerial responsibilities (Highsmith, 2002).

Yet, empirical studies reveal that engineering managers remain a fixture in Agile organizations of all sizes, from startups to global enterprises (Neto, 2023). Their continued presence raises fundamental questions about the practical realities and human factors that compel organizations to retain managerial roles despite Agile's call for minimal hierarchy. This paper investigates why engineering managers persist in Agile settings by integrating theoretical, empirical, and applied perspectives. We propose a reconciliation model that aligns managerial functions with Agile values, ensuring that management supports rather than undermines team autonomy.

## 2. LITERATURE REVIEW

### 2.1. Systematic Literature Review Methodology

We conducted a systematic literature review across IEEE Xplore, ACM Digital Library, ScienceDirect, and Google Scholar, covering publications from 2000 to 2024. Search strings combined keywords such as "Agile software development," "engineering management," "self-organizing teams," "servant leadership," and "scaled Agile."

The inclusion criteria were peer-reviewed articles, empirical studies, and systematic literature reviews that addressed management roles in Agile contexts. Exclusion criteria included non-English publications, keynote slides, and vendor-sponsored white papers without peer review.

From an initial pool of 312 records, 76 met the inclusion criteria. We coded studies for themes: managerial functions, organisational scale, leadership style, and empirical outcomes. Quantitative results were tabulated; qualitative insights were synthesised narratively.

### 2.2. Research Landscape

The inception of Agile software development marked a deliberate shift away from traditional plan-driven approaches, advocating for self-organizing teams with minimal hierarchical oversight. The Agile Manifesto emphasized "individuals and interactions over processes and tools," suggesting that roles such as Scrum Master and Product Owner could supplant traditional engineering management (Beck et al., 2001; Highsmith, 2002). Early practitioners argued that this redistribution of responsibilities rendered middle management obsolete, particularly within small, co-located teams (Cockburn & Highsmith, 2001). Empirical reviews later confirmed significant gains in responsiveness and reduced bureaucracy when management layers were flattened (Dybå & Dingsøyr, 2008). Yet these studies primarily reflected environments unencumbered by scale or regulatory complexity.

Subsequent scholarship questioned the assumption that Agile naturally eliminates managerial functions, highlighting gaps in strategic alignment, resource negotiation, and learning cultures. Edmondson (1999) demonstrated that psychological safety—a key determinant of team learning—often depends on leadership behaviors that extend beyond typical Agile roles. Similarly, Greenleaf's (1977) servant leadership model provided a theoretical foundation for managers who facilitate rather than command, aligning closely with Agile values. Moe, Dingsøyr, and Dybå (2010) further argued that organizations retain oversight mechanisms to manage risk, budget, and cross-team coordination, functions that autonomous teams alone struggle to fulfill.

Investigations into Agile transitions revealed significant role ambiguity for engineering managers as they navigated shifting expectations. Hoda, Noble, and Marshall (2013) documented how managers oscillated between directive and supportive behaviors, often defaulting to boundary-spanning activities to translate stakeholder needs into team priorities. Hoda and Murugesan (2016) identified multi-level project management challenges that required managerial intervention in planning and integration. More recent work by Shastri, Hoda, and Amor (2017, 2021) delineated distinct managerial roles—ranging from process steward to strategic coach—underscoring how managers adapt to support team autonomy while ensuring organizational accountability.

As Agile scaled beyond individual teams, frameworks such as Spotify's Tribe model, SAFe, and LeSS reincorporated management layers to handle program increments, portfolio governance, and compliance (Kniberg & Ivarsson, 2012; Leffingwell, 2016; Larman & Vodde, 2016). Empirical evidence suggests that engineering managers in these contexts serve as Tribe Leads or Release Train Engineers, coordinating dependencies, reducing integration defects, and aligning cross-functional efforts (Neto, 2023). These roles operate less as command centers and more as facilitators of alignment, ensuring agility at scale without sacrificing strategic coherence or regulatory adherence.

Recent empirical studies have deepened our understanding of how managers contribute to Agile success in practice. Luong, Jamieson, and van de Ven (2023) linked leadership coaching behaviors to enhanced psychological safety and team innovation. Sarpiri and Gandomani (2017) found that agile managers accelerate development cycles by removing impediments and promoting knowledge sharing. Itzik and Gelbard (2023) demonstrated that managerial involvement in procurement and compliance helps fill critical gaps in projects that are unsuited to pure Agile methods. Sheuly's (2013) review of startup environments further highlighted managers' roles in supporting continuous integration and communication with stakeholders, cementing the notion that modern engineering managers adapt their functions to reinforce, rather than constrain, Agile values. Five main themes emerge from the literature survey as summarized in Table 1.

Table 1.  Literature Review Summary.

| Theme | Summary of Insights | Key References |
|---|---|---|
| Challenging Early Assumptions | Foundational Agile literature assumed managers would become obsolete due to role redistribution (Scrum Master, Product Owner, Self-Managed), but this only happened in small, collocated teams of senior developers. The early dismissal of engineering managers was context-dependent and overly simplistic. | Beck et al. (2001), Highsmith (2002), Cockburn & Highsmith (2001), Dybå & Dingsøyr (2008) |
| Role Ambiguity and Evolution | Managers adapted by taking on strategic, coaching, and coordination responsibilities often unaddressed by Agile frameworks. | Hoda et al. (2013), Hoda & Murugesan (2016), Shastri et al. (2017, 2021) |
| Psychological Safety and Leadership Fit | Agile success hinges on leadership behaviors such as trust-building and coaching, underscoring managers' importance to team climate and learning. | Edmondson (1999), Greenleaf (1977), Luong et al. (2023), Moe et al. (2010) |
| Scaling and Organizational Complexity | Agile scaling and team distribution frameworks reintroduce management layers to handle program increments, inter-team dependencies, and regulatory governance. | Kniberg & Ivarsson (2012), Leffingwell (2016), Larman & Vodde (2016), Neto (2023) |
| Bridging Theory with Practice | Modern managers act as facilitators and technical bridges; filling gaps in procurement, compliance, and integration. These activities are still ignored by pure Agile theory. | Itzik & Gelbard (2023), Sheuly (2013), Sarpiri & Gandomani (2017), Campanelli & Parreiras (2015) |

By synthesizing foundational Agile principles with contemporary empirical findings, this paper reframes the enduring role of engineering managers not as relics of hierarchical control but as essential drivers of Agile practice. While early Agile literature positioned managers as obsolete (Beck et al. (2001) Highsmith (2002)), our analysis demonstrates that engineering managers perform boundary-spanning and servant-leadership functions—such as coaching teams, ensuring psychological safety, and translating strategic priorities—that are indispensable for sustained team learning and innovation (Edmondson (1999) Luong et al. (2023)). In doing so, we bridge the gap between small-team case studies and large-scale Agile implementations by illustrating how frameworks like SAFe and LeSS have reincorporated managerial layers to manage complexity, regulatory compliance, and multi-team dependencies (Kniberg & Ivarsson (2012) Leffingwell (2016)). Moreover, recent studies have shown that most Agile teams are still far from being self-managed, especially in large organizations with deep hierarchies. Kohnová and Salajová (2021) Spiegler et al. (2021) Khanagha et al. (2022).

This paper contributes a testable, three-dimensional framework, comprising contextual fit, functional necessity, and Agile compatibility, to evaluate the type of leadership needed in an Agile team. This study aims to address the long-standing debate on the role of an engineering manager in Agile software development.

## 3. BACKGROUND

### 3.1. History

The Waterfall model, which dominated from the 1970s through the 1990s, emphasized sequential phases—requirements, design, implementation, verification, and maintenance—often resulting in the late discovery of defects and poor responsiveness to changing customer needs (Boone & Khorsand, 1999). Dissatisfaction with rigid gate reviews and heavy documentation spurred calls for more adaptive approaches (Dybå & Dingsøyr, 2008).

In 2001, seventeen software practitioners authored the Agile Manifesto, which codified principles prioritizing working software, customer collaboration, and responsiveness to change (Beck et al., 2001). Agile frameworks such as Scrum redistributed managerial tasks: the Product Owner owned the backlog, the Scrum Master facilitated ceremonies, and development teams self-organized around deliverables (Schwaber & Sutherland, 2020).

Highsmith (2002) argued that the democratization of technical decision-making and shared ownership of process improvement rendered traditional command-and-control management obsolete. Early case reports described the elimination of engineering managers, with teams reporting directly to business stakeholders (Cockburn & Highsmith, 2001). However, these accounts often came from small, co-located teams with minimal regulatory constraints.

Subsequent scholarship challenged the notion that Agile naturally obviates management. Moe, Dingsøyr, and Dybå (2010) highlighted persistent needs for strategic alignment, resource allocation, and career development—functions that self-organizing teams alone struggled to fulfill. Similarly, Edmondson's (1999) work on psychological safety underscored the role of leadership in fostering trust and learning behaviors, even among autonomous teams.

### 3.2. Agile Principles and Traditional Management

Agile champions team autonomy, yet organizations routinely establish governance structures to manage risk, budgets, and compliance (Moe et al., 2010). Engineering managers often bridge this gap, providing oversight without direct task control—an uneasy balance between trust and accountability. Hoda, Noble, and Marshall (2013) observed that managers in Agile transformations face ambiguous role definitions: being too directive stifles growth, while being too hands-off leaves teams feeling unsupported. This ambiguity gives rise to "boundary-spanning" behaviors, where engineering managers leverage their technical knowledge to translate stakeholder requirements into team priorities. Edmondson's (1999) concept of psychological safety remains central to Agile success. Luong, Jamieson, and van de Ven (2023) empirically linked managerial coaching behaviors to elevated psychological safety and innovation in high-uncertainty projects.

### 3.3. Organizational Realities Driving Manager Retention

As organizations scale beyond a handful of teams, coordination complexity explodes. Frameworks like SAFe, LeSS, and Disciplined Agile Delivery reintroduce management layers for portfolio governance, program increment planning, and compliance auditing (Leffingwell, 2016; Larman & Vodde, 2016). Industries such as finance, healthcare, and aerospace operate under stringent regulatory regimes. Engineering managers ensure traceability, audit readiness, and alignment with quality standards, tasks that purely self-organizing teams may underprioritize. Agile emphasizes cross-functional capabilities, but professionals still seek career trajectories and performance evaluations. Engineering managers design development plans, facilitate mentorship, and negotiate promotions—roles vital for retention in competitive labor markets (Hoda & Murugesan, 2016).

### 3.4. Empirical Evidence

Spotify's scaled Agile "Tribe" structure features Tribe Leads (engineering managers) who coach squads while safeguarding technical health and fostering collaboration across squads (Kniberg & Ivarsson, 2012). Survey data showed that these leads improved alignment and reduced duplication of work.

Behrens, Johnson, and Li (2021) examined three global banks transitioning to Agile. Engineering managers served as program managers for Agile Release Trains, managing dependencies and coordinating regulatory submissions. Their presence correlated with a 25% reduction in integration defects.

Sheuly's (2013) systematic review of Agile in startups revealed that engineering managers often assumed technical coaching roles, supporting continuous integration pipelines and oversaw remote teams—functions critical to time-to-market acceleration.

### 3.5. Evolving Managerial Roles in Agile Contexts

Greenleaf's (1977) servant leadership philosophy aligns closely with Agile values. Managers now prioritize removing impediments, facilitating team decisions, and advocating for resources—stepping inward only when strategic intervention is required.

Campanelli and Parreiras (2015) noted that Agile teams excel at short-term delivery but may lose sight of long-term product vision. Engineering managers translate organizational strategy into product roadmaps, ensuring incremental work aligns with broader objectives.

Engineering managers thus act as effective boundary spanners and communication bridges between Agile teams and executive stakeholders. They negotiate scope changes, secure funding, and communicate progress in business terms—freeing teams to focus on technical execution (Hoda et al., 2013).

Figure 1 shows the EM transformation from traditional command-and-control management to an Agile-aligned servant leadership style across five key dimensions:

- From task assignment to team facilitation
- From performance monitoring to psychological safety
- From centralized decisions to coaching/mentoring
- From resource control to impediment removal
- From status reporting to strategic alignment

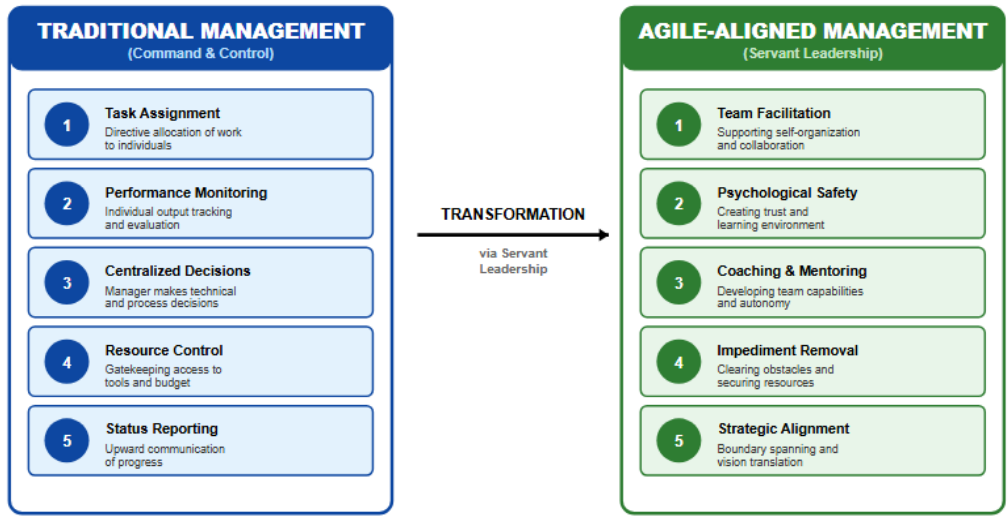

Figure 1.  Evolution of Engineering Manager Functions.

## 3.6. Findings from Research and Empirical Practice

There is consensus among the Agile experts cited in the literature that the Engineering Manager plays a pivotal role in Agile software development. Several developments in Agile practice contribute to this finding.

Firstly, the technical domain expertise of the Engineering Manager (EM) is not usually found in the Scrum Master or Project Manager role. A Scrum Master is a part-time servant leader role that can only provide initial Agile coaching and remove impediments for the Sprint team. Both functions can be performed just as readily by the EM. Technical expertise also makes the EM the best choice for serving as a conduit between the Sprint team and other managers within the organization.

Secondly, EM has formal authority and responsibilities for the career growth of the Sprint team members. The association between an EM and the Sprint team is therefore not temporary. EM is in a better position to build enduring relationships with the team than the Scrum Master or Project Manager.

Thirdly, the final goal of all Sprint teams should be to become self-managed. Yet most Sprint teams are not capable of being self-managed without proper training and coaching. With a unique techno-managerial skill set, the EM is the ideal role to facilitate this gradual transfer of power. While EMs in large, hierarchical organizations may try to subvert this leadership transfer, this can be tackled by providing the right incentives to the EMs. This responsibility lies with the Senior Leadership, working closely with the human resources leader.

Figure 2 shows the broader contextual fit of the EM compared to the Scrum Master and Project Manager.

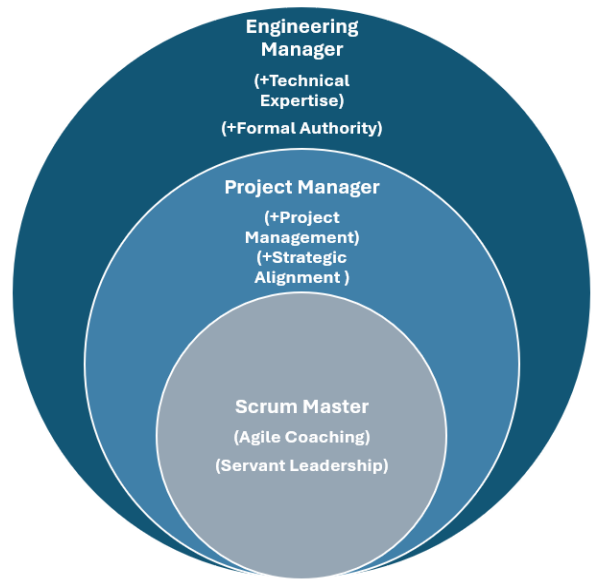

Figure 2. Role Fit for Agile Scrum Team Leadership.

## 4. PROPOSED THREE-STAGE POWER TRANSFER PROCESS

We propose a three-stage power transfer process from the EM to the Sprint Team with suggested timeframes for each stage as shown in Table 2.

Table 2. Power Transfer Process from EM to Sprint Team.

| Stage | Agile Team Leader | EM Leadership Style | Suggested Timeframe (Spiegler et al., 2021) |
|---|---|---|---|
| I. Beginning | Engineering Manager | Autocratic (Director) | First 5-10 Sprints |
| II. Transition | Engineering Manager / Senior Team Member | Democratic (Coach + Observer) | Next 5 Sprints |
| III. Self-Managed | Team Members | Consultant | Next 5 Sprints |

Figure 3 illustrates the three-stage power transition as follows:

- **Stage I (Beginning)**: Engineering Manager as autocratic director (5-10 sprints)
- **Stage II (Transition)**: Democratic coach with emerging team leaders (5 sprints)
- **Stage III (Self-Managed)**: Consultant role with autonomous team (5 sprints)

The figure includes a power transfer continuum and timeline to show the progression of team autonomy (Spiegler et al., 2021).

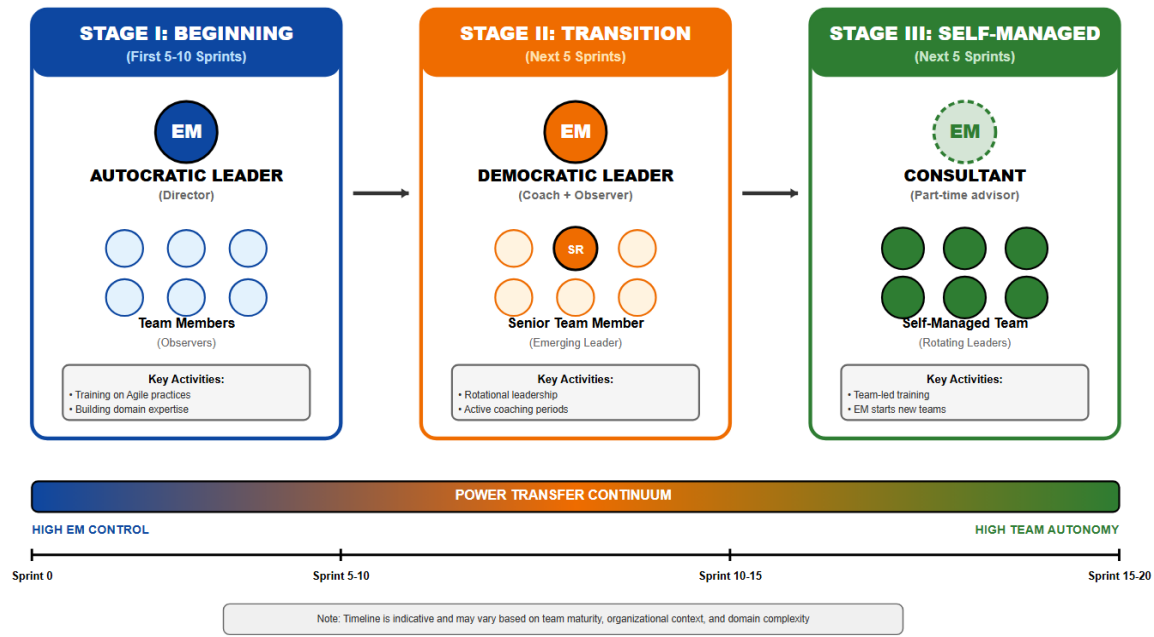

Figure 3.  Three-Stage Power Transfer Process.

In the first stage, the EM is an autocratic leader who trains the team on Agile software development best practices and builds technical domain expertise. The EM serves as a leadership role model to the team, demonstrating how to perform the activities of an Agile team leader. Team members observe and discuss regularly on the meaning of the role at the Sprint retrospectives. The team gradually builds a shared mental model of the role which leads to role clarity. This stage is expected to last for 5-10 Sprints.

In the second stage, the EM becomes a democratic leader who weaves short periods of backseat observation with active training and coaching of the Sprint team. This is a crucial stage where EMs need to ensure other managers in the organization don't take over the Agile team leader role. At the same time, if the team see that the EM is reluctant to hand over power, they must be able to report this situation directly to the human resource leader or senior leadership. In the event, a team member (usually a Senior Developer) claims the leadership role, other team members should allow the respective team member to take over that role and recognize that it is a rotational responsibility within the team. This stage is expected to last for the next 5 Sprints.

In the third stage, at least some, if not all, team members assume the role of Agile team leader. EM is in a purely consulting role on a part-time basis. EM is using the remaining time to kick off another Sprint team build. At the end of this stage, the Sprint team can be considered as Self-Managed. Any new team member will be trained and coached by the Team from now on. This stage is expected to last for the next 5 Sprints.

This framework serves as both an analytic lens and an implementation guide, enabling organizations to formally recognize the transformative role of the EM that reinforces, rather than contradicts, Agile values.

## 5. IMPLICATIONS AND LIMITATIONS

### 5.1. Implications for Practice

Organizations should rebrand engineering managers as Agile leaders, emphasizing servant-leadership competencies and boundary-spanning capacities. Clear role descriptions and behavioural expectations reduce ambiguity and align managerial activities with team autonomy.

Investing in leadership training focused on Agile–aligned behaviours—psychological safety, active listening, and strategic facilitation—bolsters managerial effectiveness. Internal coaching certifications and peer-mentoring communities support continuous skill refinement.

Project Portfolio Management (PPM) tools should incorporate features that enable managers to monitor cross-team dependencies, track strategic metrics, and surface coaching opportunities without micromanaging backlogs. Dashboards combining technical KPIs with leadership health indicators can promote balanced oversight.

### 5.2. Implications for Research

Future empirical studies should examine which leadership styles—transformational, servant, democratic—best correlate with Agile team performance, satisfaction, and innovation. Quantitative metrics (cycle time, defect rates) paired with qualitative assessments (team climate surveys) will yield robust insights.

Exploring psychological fit—the congruence between individual needs and organizational culture—in Agile contexts can reveal how engineering managers influence job satisfaction, burnout, and employee retention. Longitudinal studies tracking manager-team dyads are particularly promising.

Researchers can analyse how embedding managerial support features in Agile tools affects team autonomy and delivery outcomes. Controlled experiments comparing teams with and without tool-based managerial visibility could clarify best practices.

### 5.3. Limitations and Future Directions

This paper relied primarily on secondary data from published studies. Although systematic, our literature review may have omitted unpublished industry reports and non-English publications, which could potentially skew geographic and sectoral representation. Empirical study analyses are illustrative rather than exhaustive; further field research in diverse organizational contexts is needed. Finally, our conceptual framework requires empirical validation through large-scale surveys and controlled experiments.

**Geographic Scope Limitations:** Our literature review reveals a pronounced geographic bias toward North American and European contexts, with limited representation from Asia-Pacific, Latin American, and African markets. This Western-centric perspective may not adequately capture cultural variations in leadership styles, hierarchical expectations, or team dynamics that influence managerial roles in Agile implementations. For instance, high power-distance cultures may necessitate different approaches to the power transfer process we propose, and the timeline for achieving self-managed teams may vary significantly across cultural contexts.

**Industrial Scope Limitations:** While our analysis includes examples from finance, healthcare, and technology sectors, it underrepresents industries with unique regulatory constraints such as aerospace, defense, pharmaceuticals, and critical infrastructure. These sectors often operate under compliance frameworks that may fundamentally alter the feasibility of pure self-management. Additionally, our findings may not generalize to industries with different innovation cycles, such as manufacturing or construction, where Agile adoption patterns and managerial needs differ substantially from software development contexts.

**Organizational Scope Limitations:** The studies reviewed predominantly feature medium to large enterprises, with limited representation of micro-enterprises (fewer than 10 employees) and government agencies. Startup environments, where role fluidity is often higher and formal management structures more nascent, may exhibit different patterns of managerial evolution. Conversely, public sector organizations with entrenched bureaucratic structures may face unique challenges in implementing the three-stage power transfer process. Furthermore, our framework assumes relatively stable team compositions, which may not reflect the reality of organizations with high contractor usage, frequent reorganizations, or project-based staffing models.

**Methodological Limitations**: The three-stage power transfer timeline (15-20 sprints total) represents an idealized progression that may not account for variations in team maturity, differences in technical complexity, or resistance to organizational change. The framework also assumes continuous team membership and stable organizational support, conditions that may be unrealistic in dynamic business environments.

**Future Research Directions:** To address these limitations, we recommend: (1) comparative studies across diverse geographic regions to identify cultural moderators of Agile management practices; (2) sector-specific investigations in highly regulated industries to understand compliance-driven variations in managerial roles; (3) longitudinal studies tracking the same teams through multiple organizational transformations; and (4) examination of hybrid organizational forms, such as platform economies and distributed autonomous organizations, where traditional managerial concepts may require fundamental reconceptualization.

## 6. CONCLUSIONS

The enduring presence of engineering managers in Agile software development reflects a pragmatic reconciliation between ideological purity and organizational complexity. Far from being relics of top-down control, modern engineering managers perform vital functions, including servant leadership, boundary spanning, strategic alignment, and talent development, that complement Agile principles. By adopting Agile-compatible behaviours and leveraging PPM tools enriched with managerial insights, organizations can harness the best of both worlds: empowered, self-organizing teams guided by supportive and visionary leadership. Implementing the tri-dimensional framework outlined here will help practitioners and researchers alike navigate the evolving interplay between management and agility.

## AUTHOR

**Dr. Ravi Kalluri** received his Ph.D. in Engineering Management from the Old Dominion University, MBA in Operations and Strategy Management from Northwestern University (Kellogg), M.S. in Electrical Engineering from Stanford University, M.S. in Mechanical Engineering from Ohio State University and B.E. with Honors in Mechanical Engineering from Anna University, Chennai , India. He is presently an Assistant Teaching Professor in the MS Project Management program at Northeastern University, College of Professional Studies. His research centers on Agile Team Dynamics, and Agile Leadership.

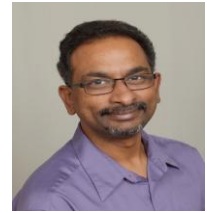